# Stanene: Atomically Thick Free-standing Layer of 2D Hexagonal Tin

*Sumit Saxena[1], Raghvendra Pratap Choudhary, and Shobha Shukla[1*]*
[1]Nanostructure Engineering and Modeling Laboratory, Department of Metallurgical Engineering and Materials Science, Indian Institute of Technology Bombay, Mumbai, MH, India 400076
[1]Contributed equally
Corresponding author: [*] sshukla@iitb.ac.in

Two dimensional (2D) layered materials have recently gained renewed interest due to their exotic electronic properties along with high specific surface area. The prospects of exploiting these properties in sensing, catalysis, energy storage, protective coatings and electrochromism have witnessed a paradigm shift towards the exploration of these sophisticated 2D materials. The exemplary performance of graphene (*1*) which is among the first of these elemental 2D materials have initiated a runaway effect in the pursuit of studying alternative 2D materials. Even though graphene has tunable exotic electronic properties (*2*), the spin-orbit (SO) coupling is weak (*3*) limiting its applications as spin filters, topological insulators etc. Topological insulators by their very nature force the electrons to travel on the surface at very high speeds thereby finding useful applications in electronic and photonic devices. Exploration of group IV elements using first principles calculations have revealed that the SO coupling increases as the atomic weight of the basis atoms in the honeycomb lattice (*4*). Tin is one of the heaviest elements in this series having strong spin-orbit coupling making it a promising applicant for room temperature topological insulator. Thus there is an urgent need to discover novel 2D materials in the post graphene age to overcome its deficiencies. We have synthesized and investigated the optical transitions in 2D material referred to as few-layer stanene (FLS). These are analogous to few-layer graphene and can be visualized by replacing carbon atoms by tin on a graphene lattice. Stanene by itself is presumed not to exist naturally. We have been able to synthesize from mono to few atomic layers of free standing stanene and characterize them optically using UV-Vis absorption, Raman and photoluminescence spectroscopy. First principles calculations have been performed to interpret experimental results.

Ultra-fast light matter interaction in liquid ambience followed by hydrazine treatment, has been used to synthesize high quality stanene sheets and identified using HRTEM, SAED and EDAX analysis. The HRTEM image of a representative sample (Fig 1) shows hexagonal arrangement of atoms. This was verified using SAED pattern. The inter-planar spacing 'd' calculated using the SAED pattern was found to be $d_1 \sim 0.25$ nm and $d_2 \sim 0.16$ nm. These are in good agreement with the X-ray diffraction pattern calculated from the optimized structure obtained from first principles calculations. These planes are parallel to the zigzag and armchair edges in the stanene lattice respectively. The chemical analysis using EDAX shows the peaks corresponding to Sn around ~4KeV.

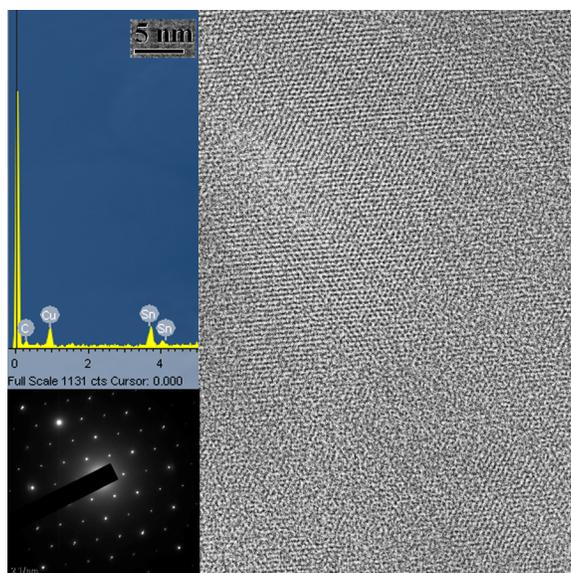

Figure 1:- Large area HRTEM image of hexagonal stanene lattice. Top left inset shows the elemental composition using EDAX. Carbon and copper peaks arises due to the TEM grid used. Bottom left inset shows electron diffraction pattern obtained from the sample showing hexagonal lattice.

The as prepared stanene samples show strong absorption peak ~196 nm (6.325 eV) and two shoulders at ~225 (5.510 eV) and ~250 nm (4.959 eV) (Fig 2a). The samples were reduced using hydrazine which resulted in disappearance of the shoulder at ~225 nm. However this shoulder reappears again on oxidation of the reduced samples as also in the case of graphene (*5*) samples. This suggests that the shoulder ~225nm appears due to the transitions originating from the oxidation of the stanene samples. The photoluminescence spectra (Fig 2b) show strong emission peaks at ~332 (3.734eV), 339nm (3.657 eV) along with small emission peaks at 402 (3.084eV), 423 (2.931eV), 494 (2.509 eV) and 524nm (2.366 eV). The strong peak at ~461nm corresponds to hexane in the sample and marked by '*' in the photoluminescence spectrum. No significant difference is observed in the emission spectrum of the reduced and unreduced samples. The evaluation of these transitions is very important for interpretation of optical properties. Raman studies using 532 nm Laser beam suggest the presence of a very strong peak ~149cm$^{-1}$ (Fig 3b) and is in good agreement to placement of gamma phonon in buckled stanene structures in the phonon dispersion. Prolonged exposure of high intensity laser beam results in shifting of this peak ~167cm$^{-1}$. This shift of the peak is understood to occur due to change in oxidation state of the Sn atoms. The appearance of additional vibrational modes at ~288 cm$^{-1}$ suggesting that large does of high intensity laser beam oxidizes stanene layers forming domains of SnO$_x$ in stanene layers. The peaks ~288 cm$^{-1}$ are in good agreement with the E$_u$ mode observed in the IR spectrum of crystalline SnO$_2$(*6*).

First principles calculations using density functional theory as implemented in VASP(*7*) suggest that stanene stabilizes in low buckled configuration with buckling height of Δ=0.85Å. This is in good agreement with previously reported data (*8*). The buckling results in overlap between the π and σ states. The linear energy-momentum dispersion coupled with large Fermi velocity reveal the presence of Dirac cone in the electronic band structure of stanene at the K point. Presence of saddle point at the M point along the high symmetry direction (Fig 3a) is also observed. It is however important to note that inclusion of spin-orbit coupling introduces a gap of ~ 73 meV (*9*). The bandgap thus arising at the Dirac point facilitates the 2D material transition from a semi-metallic to a quantum spin Hall insulator (*4*). The electrons moving along the M-K (M-Γ) direction are expected to display negative effective mass. The occurrence of parallel bands along the Γ-M-K-Γ directions in the band structure is of peculiar interest because of their potential enhanced excitonic effects. The presence of parallel dispersions is understood to facilitate the excited electron and hole to move together giving rise to enhanced e-h pairs. Similar features have been understood to explain effects not only in semiconductors such as silicon (*10*) but also semi-metals as graphene (*11*). The symmetric peak profile along with emission wavelengths suggests that the probable origin of the peaks at 332, 339, 402 and 423 nm can be associated with excitonic effects in stanene due to the transition of electrons from $\pi$ to $\pi^*$ bands in stanene. The lower emission peaks at 494nm and 524 nm are understood to arise from direct optical transition between the highest valence and lowest conduction band (Fig 3a).

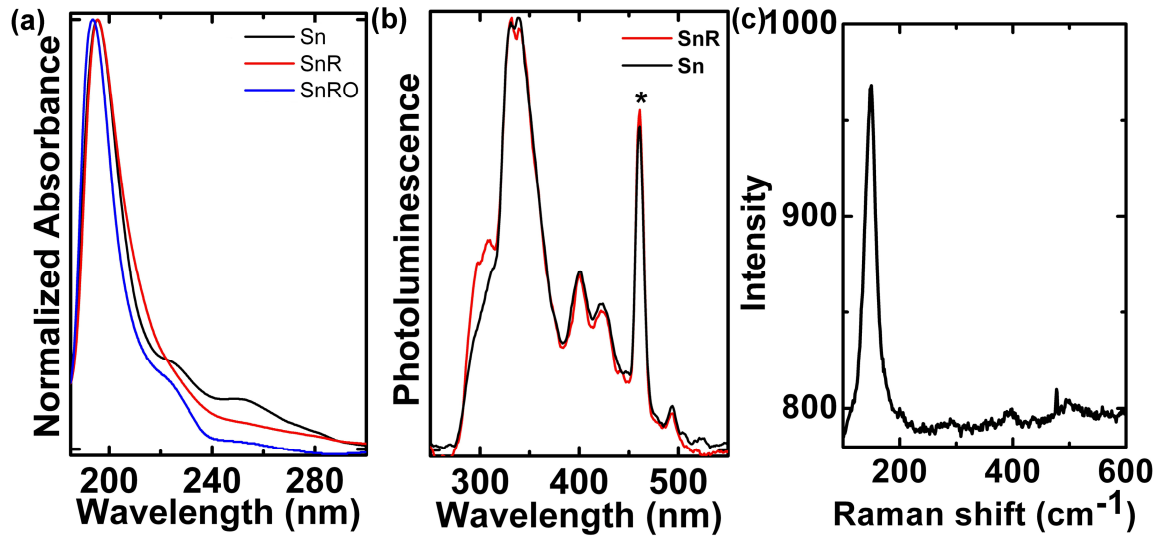

Fig 2(a) Normalized UV spectrum of as prepared (Sn) stanene sheets, stanene sheets reduced using hydrazine (SnR) and subsequently oxidized using $H_2O_2$ (SnRO). The shoulder ~225 nm arises due to oxidation. (b) Normalized photoluminescence spectrum of as prepared stanene (Sn) and reduced samples (SnR). (c) Micro-Raman spectrum of as prepared Sn sample using 532 nm Laser.

The normalized partial density of states shows that the electronic states near the Dirac point are dominated by $\pi$ states with major contribution coming from the out of plane $p_z$ orbital (Fig 3b). This suggests that most of the low energy optical transitions occur due to the transitions between the $\pi$ and $\pi^*$ states. The $\sigma$ states lie farther down below the $\pi$ states about 3 eV below the Fermi level. The optical properties of the material can be described in terms of complex dielectric function $\varepsilon = \varepsilon_1 + i\varepsilon_2$ and the absorption coefficient is directly proportional to imaginary part of the complex dielectric function ($\varepsilon_2$). $\varepsilon_2$ is obtained using first principles density functional calculations within the generalized gradient approach. Stanene displays pronounced asymmetric peak in the UV region at photon energy of ~6.325eV. This is understood to arise from the band to band transition near the saddle point singularity at the M point. The peaks at ~191 nm and ~233nm in the calculated $\varepsilon_2$ (Fig 3c) shows a remarkable similarity with the experimental data (Fig 2a)

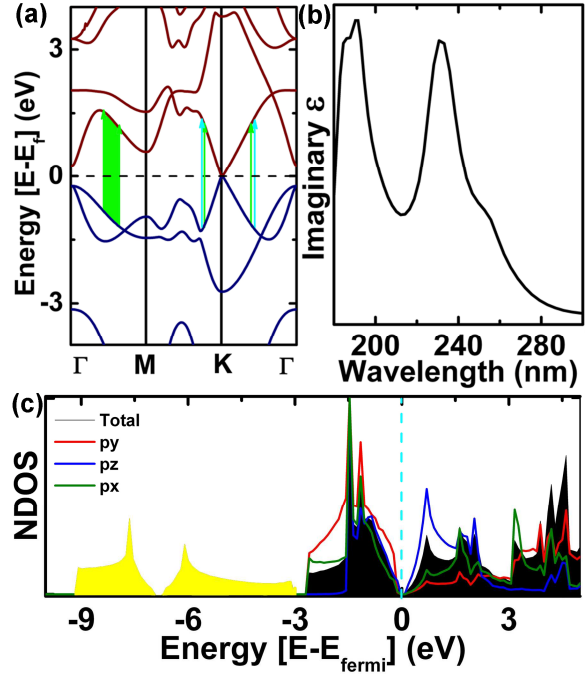

Fig 3: Electronic structure of stanene obtained using ab-initio methods. (a) Electronic band structure of stanene. The emission peak at ~524 nm and ~494 nm are marked by probable green and cyan transitions. (b) Normalized partial density of states of stanene. The yellow shaded PDOS (-9 to -3 eV) in valence band region is due to contribution from σ states. The vertical dashed line at 0eV (cyan) represents Fermi level. (c) X component of the imaginary part of the dielectric constant.

The availability of UV excitons in stanene makes it attractive for optoelectronic applications. The opening of band gap due to spin orbit coupling makes it promising candidate for room temperature topological insulator thus making stanene essentially an all surface 2D material.

**Acknowledgements**
This work was supported by the Department of Science and Technology, Solar Energy Research Initiative (SERI), Government of India grant via sanction order no. DST/TM/SERI/2k10/12/(G) and the Industrial Research and Consultancy Services, Indian Institute of Technology Bombay, grant no. (11IRCCSG025)